\documentclass[sigconf,authorversion]{acmart}

\usepackage[normalem]{ulem}

\AtBeginDocument{%
  \providecommand\BibTeX{{%
    \normalfont B\kern-0.5em{\scshape i\kern-0.25em b}\kern-0.8em\TeX}}}

\copyrightyear{2024} 
\acmYear{2024} 
\setcopyright{rightsretained} 
\acmConference[WSDM '24]{Proceedings of the 17th ACM International Conference on Web Search and Data Mining}{March 4--8, 2024}{Merida, Mexico}
\acmBooktitle{Proceedings of the 17th ACM International Conference on Web Search and Data Mining (WSDM '24), March 4--8, 2024, Merida, Mexico}\acmDOI{10.1145/3616855.3635730}
\acmISBN{979-8-4007-0371-3/24/03}

\usepackage{booktabs} 
\usepackage{color}
\usepackage[inline]{enumitem}

\begin{document}

\fancyhead{}
\title{Leveraging User Simulation to Develop and Evaluate Conversational Information Access Agents}

\author{Nolwenn Bernard}
\affiliation{%
  \institution{University of Stavanger}
  \city{Stavanger}
  \country{Norway}
}
\email{nolwenn.m.bernard@uis.no}

\begin{abstract}
We observe a change in the way users access information, that is, the rise of conversational information access (CIA) agents.
However, the automatic evaluation of these agents remains an open challenge. 
Moreover, the training of CIA agents is cumbersome as it mostly relies on conversational corpora, expert knowledge, and reinforcement learning.
User simulation has been identified as a promising solution to tackle automatic evaluation and has been previously used in reinforcement learning.
In this research, we investigate how user simulation can be leveraged in the context of CIA.
We organize the work in three parts. We begin with the identification of requirements for user simulators for training and evaluating CIA agents and compare existing types of simulator regarding these. Then, we plan to combine these different types of simulators into a new hybrid simulator. Finally, we aim to extend simulators to handle more complex information seeking scenarios.  
\end{abstract}



\maketitle

\section{Introduction}

Conversational Information Access (CIA) is a relatively new field, yet it attracts a growing attention from the research community and industry especially with the rise of conversational agents based on large language models (LLMs).
Conversational information access agents support multiple conversational goals that are related to complex information seeking, exploratory information gathering, and recommendation~\citep{Culpepper:2018:SIGIRForum}.
Conversational systems for search, question answering, and recommendation are categorized as CIA agents. 
This field is at the intersection of multiple domains such as dialogue systems, information retrieval, and computer human interaction, while having some distinct characteristics. 
One of the main challenge in the field regards the evaluation of CIA agents.
As CIA agents are becoming more and more popular to users, it is important to ensure their quality before releasing them to avoid negative, and potentially harmful, user experiences; this is where user simulation may play a major role. Indeed, it has been identified as a simple, cost and time efficient solution to evaluate these agents offline~\citep{Balog:2021:DESIRES}.
It is a complement to online evaluation and user studies that are performed with humans, hence, expensive, difficult to reproduce and scale~\citep{Balog:2023:arXiv}.
Furthermore, user simulation has been used in the context of reinforcement learning to train conversational agents or a specific component such as dialogue management~\citep{Schatzmann:2006:KER,Tseng:2021:ACL}. 

While user simulators tackle some limitations of online evaluation, they remain at their infancy and have limitations~\citep{Balog:2023:arXiv}.
Moreover, we observe that user simulators tend to focus on a single conversational goal (recommendation, search, or question answering). However, CIA also supports scenarios mixing multiple conversational goals, e.g., search on benefits of running before asking a recommendation for a pair of running shoes.
Therefore, there are opportunities to study simulators' limitations in the context of CIA and to extend them to support more complex multi-goal scenarios.
Our main objective is to study how user simulation can be leveraged in the context of CIA. We note that we will consider the two common uses mentioned before: training and evaluation. 
Our research is organized in three subsequent parts with their associated research questions (\emph{RQs}):
\begin{enumerate}
    \item The tasks of training and evaluation of a CIA agent have different objectives, thus, simulators may have different requirements and desiderata. For example, interpretability is more important for evaluation than training, as it gives insights for the analysis of the results that is not necessary for training.
    We would like to compare different types of simulator with regards to these requirements. In this part, we will answer \emph{RQ1}: What are the requirements for user simulators when training and evaluating a single goal CIA agent? How do different types of simulator compare with regards to these requirements?
    \item The previous comparison of different types of simulator will highlight their strengths and weaknesses with regards to the task's requirements. In this part, we want to study if and how the different types of simulator can be combined into a hybrid user simulator. Thus, we will address \emph{RQ2}: How to create a hybrid user simulator taking the best of each type of user simulator? 
    \item Finally, we want to focus on the extension of existing work to support more complex scenarios mixing multiple conversational goals. It also includes to study existing datasets in CIA in terms of suitability for this task. Hence, we have \emph{RQ3}: How suited are existing datasets for CIA agents handling multiple conversational goals? How to extend current user simulators to handle scenarios mixing multiple conversational goals?
\end{enumerate}

\vspace{-0.75\baselineskip}
\section{Background}
\label{sec:related}

User simulation was introduced in the field of dialogue systems few decades ago~\citep{Eckert:1997:ASRU}, and its potential in conversational information access was recently highlighted~\citep{Balog:2021:DESIRES}.
It has been mostly used for training and evaluation of dialogue systems.
User simulation benefits from current advances in the fields of dialogue management and conversational information access. 
Indeed, traditional user simulators may be seen as simplistic and highly tailored to one application, however, these can be modified and/or extended with more robust components thanks to new resources (e.g., methods and datasets). More recently, user simulators based on large language models have been proposed to tackle some limitations of traditional approach such as the lack of diversity in the answers~\citep{Davidson:2023:arXiv}. 

We can identify different types of user simulator that can be used for various conversational goals, i.e., search, question-answering, and recommendation. 
Agenda-based simulators~\citep{Schatzmann:2007:NAACL} are widely used, especially for conversational recommendation as in~\citep{Zhang:2020:KDD}. The agenda represents the information need and the sequence of actions the user wants to take next. 
Then, there are simulators based on deep neural networks, e.g.,~\citep{Gur:2018:SLT,Lin:2021:SIGDIAL,Lin:2022:SIGDIAL}.
Last, we identify simulators based on large language models, e.g.,~\citep{Davidson:2023:arXiv,Sun:2023:TOIS,Owoicho:2023:SIGIR,Sekulic:2022:WSDM}.
Commonly, simulators are developed to interact with task-oriented dialogue systems such as CIA agents. However, we note that they are mostly used in scenarios involving a single conversational goal. 
Some examples include USi~\citep{Sekulic:2022:WSDM} and ConvSim~\citep{Owoicho:2023:SIGIR} that focus on answering clarification question in addition to feedback for the latter in the context of conversational search.
It might be explained by the fact that a majority of existing benchmarks and datasets available for CIA do not include conversations mixing conversational goals~\citep{Bernard:2023:SIGIR}.

A critical hurdle to the adoption of user simulation, especially for evaluation, is the assessment of user simulators themselves~\citep{Balog:2023:arXiv}. Indeed, current user simulators are imperfect, therefore, this should be considered when interpreting experiments and evaluation outcomes. 
Current solutions to evaluate user simulators include human evaluation, similar methods as for conversational agents (e.g., BLEU and success rate), and comparison of generated conversations with human-human conversational corpora.
Recently, \citet{Sun:2023:TOIS} proposed a tester-based evaluation framework to automatically assess user simulators. They evaluate how well a simulator can rank variants of a task-oriented conversational agent with regards to a metric compared to an expected ranking.

Conversational frameworks including off the shelf user simulators have been released to facilitate their use in CIA. 
Among these, ConvLab-3~\citep{Zhu:2022:arXiv} provides an implementation of TUS~\citep{Lin:2021:SIGDIAL} and GenTUS~\citep{Lin:2022:SIGDIAL} and, more recently, it has been extended to include LLM-based user simulators.
PyDial~\citep{Ultes:2017:ACL} comprises the implementation of the agenda-based simulator~\citep{Schatzmann:2007:NAACL}; it is important to note that the simulation is done on the semantic level, i.e., does not use natural language processing the analyse and produce an utterance.
Unlike the aforementioned frameworks, UserSimCRS~\citep{Afzali:2023:WSDM} is a toolkit tailored for the evaluation of conversational recommender systems. Furthermore, it provides novel components especially to have a better control over the simulator's behavior and preferences.

\section{Research Overview}
\label{sec:overview}

\begin{figure}
    \centering
    \includegraphics[scale=0.37]{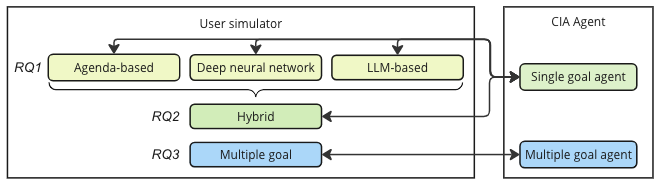}
    \caption{Conversational setting with participants associated to a research question (color-coded).}
    \label{fig:rqs}
\end{figure}

As stated above, this research aims to study how to leverage user simulation in conversational information access. 
We organize the work in three parts that are detailed below. 
In terms of setting, we always consider that two participants are involved in a conversation: a user simulator and a CIA agent.
Fig.~\ref{fig:rqs} presents an overview of the research questions and dialogue participants involved in each. 

In the first part (\emph{RQ1}), we focus on the requirements and desiderata for user simulators when training and evaluating CIA agents.
While the same user simulators can be used for both training and evaluation, we believe that different requirements and desiderata are needed for each of these tasks.
Indeed, training and evaluation have different goals, that is, to optimize a CIA agent with regards to an objective and to assess its performance in a real-world scenario, respectively.
For evaluation, it is important to have a user simulator that can mimic real users as closely as possible. 
It includes the linguistic fluency and diversity of the simulator's utterances, but also the conversational strategies that are not always optimal. 
Interpretability is desirable for evaluation as it can provide insights when analyzing the results and parameters that can control the simulated users' behavior (e.g., patience and cooperativeness).
For training, interpretability is not a necessity as we focus on optimizing the CIA agent's performance.
Furthermore, linguistic fluency and diversity are less important as the emphasis should be on the content of utterances.
For example, when training the dialogue manager of a CIA agent, we can use both natural language and structured representations of utterances.
Once the requirements and desiderata are identified, we will propose evaluation metrics to quantify them and compare existing types of user simulator using these metrics.

The aforementioned comparison of user simulator types serves to answer \emph{RQ2}.
Indeed, it is used to identify the strengths and weaknesses of each type of user simulator.
Then, the idea is to propose a hybrid approach that combines the strengths of different types of user simulator while mitigating their weaknesses.
For example, we can think of combining agenda- and LLM-based simulators, as the former is more interpretable than the latter but less flexible.  
To answer \emph{RQ2}, we need to ensure that a combination of simulator types is possible and that the resulting user simulator can meet the requirements and desiderata identified in \emph{RQ1}.

As shown in Fig.~\ref{fig:rqs}, the previous research questions considered CIA agents handling a single conversational goal.
For \emph{RQ3}, we consider CIA agents that can handle multiple conversational goals. 
However, we observe that most of the available resources in the field of CIA are tailored to a single conversational goal.
Consequently, existing agents tend to be designed for a single goal. 
Hence, we have recently introduced a novel dataset, MG-ShopDial, that contains human-human conversations mixing goals in the domain of e-commerce~\citep{Bernard:2023:SIGIR}.
We plan to use it to develop a multi-goal CIA agent that will converse with an adapted user simulator.
Building such user simulator is the answer to \emph{RQ3}.
Thanks to the knowledge acquired from \emph{RQ1} and \emph{RQ2}, we envision to extend existing user simulation to handle multiple conversational goals.

\section{Research Methodology}
\label{sec:method}

Building a user simulator for CIA is a challenging task, that may explain why previous work in the field mostly focus on single goal scenarios.
Therefore, we will investigate how user simulation can be leveraged starting with single goal scenarios and gradually moving to more complex ones.
Using this approach, we will be able to facilitate comparison with existing work and identify critical components to create a user simulator that can mimic user behavior in complex information seeking scenarios.

In practice, we will use an axiomatic analysis for the requirements of user simulation.
We use user studies and crowdsourcing for data collection and human evaluation. 
For the automatic evaluation of our user simulators, we will use and, when needed, extend the tester-based framework proposed in~\citep{Sun:2023:TOIS}.
Regarding the implementation of our user simulators and CIA agents, we want to use standard libraries and frameworks like DialogueKit and UserSimCRS~\citep{Afzali:2023:WSDM} in order to ensure their reusability and extendability.
For the experiments, we plan to use existing research platforms and benchmarks, such as IAI MovieBot~\citep{Habib:2020:CIKM} for CRS and TREC CAsT\footnote{\url{https://www.treccast.ai}} for CSS, whenever possible. 
\section{Discussion}
\label{sec:discussion}

There are several points that we would like to discuss at the Doctoral Consortium.
First, it is noteworthy that major obstacles to a wide use of user simulation for automatic evaluation of CIA agents are its trustworthiness and realism. Hence, during this Consortium, we want to ask mentors: What kind of experiments would convince them, as peers and reviewers, about the utility of simulators?

Another discussion point concerns the strategy to design and perform an efficient and meaningful human evaluation for both CIA agents and user simulators.
Human evaluation for each of these may have different purposes, for example, for user simulators it may be used to assess their quality and resemblance to human behavior, while for CIA agents it can assess their performances with regards to an objective and also their user experience.
Despite being time consuming and expensive, human evaluation is valuable to complete and support automatic evaluation.
Indeed, the interpretation and validity of automatic evaluation is limited by the quality of the user simulator. For example, if the user simulator has known limitations, then the results should be interpreted by taking them into account.
Therefore, we would like to get advice on how to effectively design a user study and what are the common errors to avoid based on mentors' experiences. 

The last point relates to the rapid development of large language models and their potential in the fields of user simulation and CIA.
Indeed, LLMs attract an increasing attention and recently are getting more popular to create conversational agents and user simulators. 
While, they have proven their efficiency in many tasks, we wonder if they should replace traditional approaches, such as agenda-based user simulation, in the long term. 
We believe that there are still some unknowns regarding their use in user simulation, for example, how much control do we have over their behavior?
Therefore, we would like to collect opinions from the mentors on the place for LLMs in user simulation.

\bibliographystyle{ACM-Reference-Format}
\bibliography{wsdm2024-dc.bib}

\end{document}